\documentclass[aip,pof,amsmath,amssymb]{revtex4-1}
\usepackage{graphicx}

\begin{document}

\newcommand{\fdrive}{f_\text{drive}}
\newcommand{\rbead}{R_\text{bead}}
\newcommand{\rdimer}{R_\text{dimer}}
\newcommand{\wi}{W \! i}

\title{Fluid Elasticity Can Enable Propulsion at Low Reynolds Number}

\author{Nathan~C.~Keim}
\email{nkeim@seas.upenn.edu}
\author{Mike~Garcia}
\author{Paulo~E.~Arratia}
\email{parratia@seas.upenn.edu}
\affiliation{Department of Mechanical Engineering and Applied Mechanics, University of Pennsylvania, Philadelphia, PA 19104}

\date{\today}

\begin{abstract}

Conventionally, a microscopic particle that performs a reciprocal stroke cannot move through its environment. This is because at small scales, the response of simple Newtonian fluids is purely viscous and flows are time-reversible. We show that by contrast, fluid elasticity enables propulsion by reciprocal forcing that is otherwise impossible. We present experiments on rigid objects actuated reciprocally in viscous fluids, demonstrating for the first time a purely elastic propulsion set by the object's shape and boundary conditions. We describe two different artificial ``swimmers'' that experimentally realize this principle.

\end{abstract}

\maketitle

A striking feature of Newtonian viscous flows is that they can be time-reversible.\cite{Happel:1983} This feature is often referred to as kinematic reversibility and has important consequences, for example, in fluid transport in micro- and nano-fluidic devices,\cite{Stone:2006cr} mixing,\cite{Ottino:1989p2366} and self-propulsion of microorganisms.\cite{Lauga:2009p3877} For microorganisms swimming in simple liquids, linear viscous stresses that scale as $\mu V/L$ are much larger than stresses from nonlinear fluid inertia, which scale as $\rho V^{2}$, where $V$ and $L$ are the characteristic velocity and length-scale, and $\rho$ and $\mu$ are the fluid density and viscosity. For these swimmers, the ratio of inertial to viscous stresses, calculated as the Reynolds number $Re = \rho VL/\mu$, is often $10^{-3}$ or smaller. Because the swimmer has a density comparable to that of the fluid, its own inertia is also negligible. The resulting kinematic reversibility means that only non-reciprocal deformations of the swimmer can break time-reversal symmetry and result in net motion; this is known as the ``scallop theorem.''\cite{Purcell:1977p8045}

However, the hydrodynamic stresses on a microorganism need not be purely viscous. Many microorganisms live in complex fluid media that contain solids and/or polymers~\cite{Lauga:2009p8056}. Fluids such as gels, mud, intestinal fluid, and human mucus are not Newtonian, and often possess viscoelastic behavior. Recent work has begun to explore the important higher-order effects of fluid elasticity on swimmers that can also move through Newtonian (non-elastic) fluid. Theoretical and numerical studies have shown that fluid elasticity can significantly affect the propulsion speed and efficiency of microorganisms,\cite{Chaudhury:1979km,Fu:2007p5591,Lauga:2007p3864,Teran:2008p3896,Lauga:2009p8056,Leshansky:2009hr,Fu:2009p3840,Teran:2010p3897,Zhu:2011p6223,Zhu:2012ht} and breaks the time-reversal symmetry between pushers and pullers.\cite{Zhu:2012ht} Controlled experiments have shown that fluid elasticity usually hinders propulsion compared to Newtonian fluid,\cite{Shen:2011p8052} although there is evidence of an increase in propulsion speed for rotating helices in highly elastic fluids.\cite{Liu:2011wk} It is becoming increasingly clear that the presence of elastic stresses in the medium can modify swimming in a nontrivial way.

The possibility, however, that fluid elasticity can \emph{enable} rather than modify propulsion, circumventing the scallop theorem, is still largely unexplored. Propulsion enabled by fluid elasticity has been predicted for three special cases of reciprocal motion: a flapping surface extending from a plane\cite{Normand:2008p5115,Pak:2010p5116}; a sphere which generates small-amplitude sinusoidal motion of fluid along its surface~\cite{Lauga:2009p8056}; and a ``wriggling'' cylinder with reciprocal forward and backward strokes at different rates.\cite{Fu:2009p3840} However, there remains no experimental demonstration, and such propulsion of free, finite-amplitude swimmers has not been studied at all.

\begin{figure}
\begin{center}
\includegraphics[width=4in]{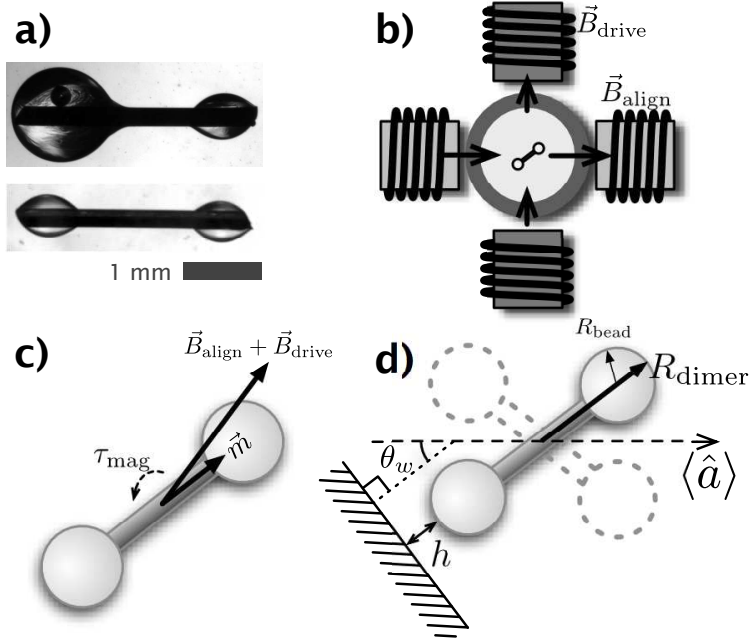}
\end{center}
\caption{\textbf{(a)} Typical ``swimmers.'' Two epoxy beads are joined by steel wire to form polar (asymmetric) and symmetric dimers.
\textbf{(b)} Top view of experiment. Two aligning electromagnets at constant current are orthogonal to two driving magnets, controlled by a computer. \textbf{(c)} The dimer with magnetization $\vec m$ experiences torque $\tau_\text{mag}$ to align with the magnetic field. \textbf{(d)} Dimer geometry. Dimer orientation $\hat a$ oscillates around $\langle \hat a \rangle$, which is parallel to $\vec B_\text{align}$. If a wall is present, the smallest separation between it and the dimer is the gap size $h$; typically 30~$\mu$m at the start of the experiment. The length of the dimer is $2\rdimer$; the bead at each end has radius $\rbead$.}
\label{fig_setup}
\end{figure}

\begin{figure}
\begin{center}
\includegraphics[width=5in]{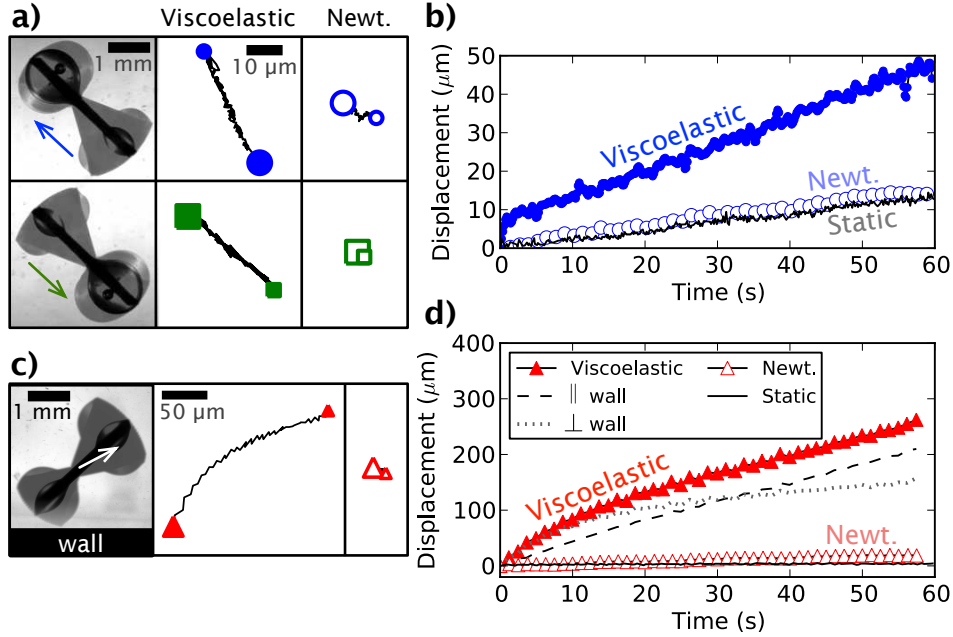}
\end{center}
\caption{(color online) Locomotion at low $Re$ due to fluid elasticity. \textbf{(a)}~Polar dimer at $De=5.7$. Left: dimer superimposed on silhouettes of motion over 1 cycle at $\fdrive = 0.4$~Hz. At top and bottom the dimer is placed in opposite orientations, but with the same actuation in each case. Middle, right: corresponding stroboscopic centroid trajectories, plotted over 50~s of driving at 2.8~Hz in viscoelastic fluid, proceeding from the large symbol to the small one. In Newtonian fluid there is negligible net translation from \emph{e.g.}~inhomogeneities in the magnetic field. The direction in viscoelastic fluid is set by the dimer shape; in Newtonian fluid, it is not. \textbf{(b)}~Net displacement \emph{vs.}\ time for upper row of (a), with same symbols. The black ``static'' line shows motion with $\vec B_\text{align}$ only (no driving).
\textbf{(c), (d)} Corresponding plots for a symmetric dimer next to a wall with $\theta_w = 45^\circ$ and $De=0.8$. As the dimer moves away from the wall, velocity perpendicular to the wall decays, while velocity parallel to the wall is constant. }
\label{fig_trajectories}
\end{figure}

In this Letter, we consider the question of whether viscoelasticity alone can enable propulsion in the absence of inertia (\emph{i.e.} low $Re$) by actuating a single rigid object reciprocally in a very viscous fluid. A propeller (``swimmer'') such as one shown in Fig.~\ref{fig_setup} is immersed in a fluid and repeatedly reoriented by a magnetic field. The effects of inertia are absent due to high fluid viscosity ($\sim 4 \times 10^4$~cSt), resulting in $Re \alt 10^{-4}$, comparable to that of a swimming microorganism. By applying only magnetic torques, our apparatus reciprocally actuates just one degree of freedom in the system, the dimer's orientation $\hat a$. For a purely viscous Newtonian fluid at low $Re$, we find no net motion because $\hat a(t)$ is cyclic. Yet when a small amount of polymer is added to the fluid, making it viscoelastic, the same ``stroke'' results in propulsion, in a direction set by the dimer's shape and boundary conditions (cf. Fig.~\ref{fig_trajectories}). While the dimers are not strictly self-propelled and so are not true swimmers, the magnetic field provides only a reciprocal torque and does not itself create or direct propulsion. This is thus the first experimental demonstration of \emph{purely elastic} propulsion, wherein fluid elastic stresses are the sole source of net motion.

Two experimental systems are used: (i) a polar (asymmetric) dimer far from any boundaries and (ii) a symmetric dimer near a wall, as shown in Fig.~\ref{fig_setup}(a). Each dimer consists of a piece of carbon steel wire of length $2\rdimer = 2.5$--3~mm and diameter 230~$\mu$m, with an epoxy bead of diameter $2\rbead \sim 500$~$\mu$m at each end. The dimer has orientation $\hat a$ and is magnetized with moment $\vec m = \hat a m$, so that a uniform magnetic field $\vec B$ reorients it with torque $\vec \tau_\text{mag} = \vec m \times \vec B$, as depicted in Fig.~\ref{fig_setup}(c). For experiments with the dimer next to a wall, a glass cover slip serves as a flat, vertical boundary (Fig.~\ref{fig_setup}[d]).

The dimer is immersed in a container (50~mm tall, 30~mm in diameter) of either Newtonian or polymeric fluid (Fig.~\ref{fig_setup}[b]). The Newtonian fluid is a 96\%-corn~syrup aqueous solution (by mass) with kinematic viscosity $\mu/\rho$ of approximately $4 \times 10^4$~cSt. The polymeric solution is made by adding 0.17\% (by mass) of high-molecular-weight polyacrylamide (PAA, $M_W = 10^6$) to a viscous Newtonian solvent (93\%-corn syrup aqueous solution). The solution is considered dilute: the overlap concentration $c^{*}$ for PAA is $\sim 0.34$\% ($c/c^* = 0.5$).\cite{Bird:1987p2262} Using a strain-controlled rheometer, we find that the PAA solution is an elastic fluid with nearly constant viscosity, varying with a power law index of $n = 0.96$ up to 50~s$^{-1}$.\cite{elmosupp} This way, the effects of shear-thinning viscosity can be decoupled from those of elasticity.\cite{Stokes:2001p3781,James:2009p3759} The fluid relaxation time $\lambda$, measured in stress relaxation tests, is approximately 2~s.\cite{elmosupp}

A schematic of the apparatus is shown in Fig.~\ref{fig_setup}(b). Four electromagnets reorient the dimer in a fluid cell. A reciprocal ``wiggling'' motion is achieved with two diametrically opposed electromagnets generating a constant field $\vec B_\text{align}$, and a pair of orthogonal magnets generating the AC field $\vec B_\text{drive}$. The driving magnet current is controlled by a computer via a power amplifier. The magnitude of $\vec B_\text{align}$ and amplitude of $\vec B_\text{drive}$ are $\mathcal{O}(10^3$~G$)$. The amplitude of dimer rotation is nearly 45$^\circ$, decaying at high $\fdrive$ in some experiments as detailed below. For polymeric fluid, we define the Deborah number $De$, the product of the longest fluid relaxation time $\lambda$ and the driving frequency $\fdrive$.

In addition to the magnetic torque, the dimer also experiences an undesirable translational force
\begin{equation}
\label{magtrans}
\vec F_\text{mag} = \vec \nabla (\vec m \cdot \vec B)
\end{equation}
due to inhomogeneities in the magnetic field. The dimer is positioned so that when both the aligning and driving fields are at full strength, the translation velocity from $\vec F_\text{mag}$ is $\leq 0.3$~$\mu$m/s. This velocity is an upper bound on the translation in our experiments that can be attributed to $\vec F_\text{mag}$: the sinusoidally-varied driving field is only briefly at full strength, and $\vec m$ and $\vec B$ are not aligned as the dimer rotates, reducing the dot product in Eq.~\ref{magtrans}.

Evidence of purely elastic propulsion is shown in Figs.~\ref{fig_trajectories}(a,~b) for the polar dimer far away from boundaries at $De=5.7$ ($\fdrive=2.8$~Hz; $Re=1.2 \times 10^{-4}$) and Fig.~\ref{fig_trajectories}(c,d) for the symmetric dimer near a wall at $De=0.80$ ($\fdrive=0.4$~Hz; $Re=6.6 \times 10^{-5}$). The dimers are imaged using a CCD camera to extract orientation and centroid position. The camera is aligned with the vertical axis and the apparent horizontal motion from sedimentation is $\alt 0.05$~$\mu$m/s. The data show a striking contrast between performing reciprocal motion in Newtonian and in viscoelastic fluid. Figure~\ref{fig_trajectories}(a,~b) shows that in viscoelastic fluid, far from any boundaries, the polar dimer is able to achieve net motion at constant speed even under reciprocal forcing; no net motion is observed in a Newtonian fluid under the same conditions. The polar dimer moves in the direction of its large end, as shown by the arrows in Fig.~\ref{fig_trajectories}(a).

Net motion is also observed for a symmetric dimer near a wall in viscoelastic fluid (Figs.~\ref{fig_trajectories}[c, d]). The symmetric dimer translates away from and along the wall with approximate direction $\langle \hat a \rangle$, as shown in Fig.~\ref{fig_trajectories}(c). Corresponding behaviors are seen in variants of the geometry where the dimer is flipped by $180^\circ$ or $\theta_w$ is varied in increments of $90^\circ$. Figure~\ref{fig_trajectories}(d) shows the typical net displacement of these trajectories as a function of time. After a short transient, the symmetric dimer immersed in a viscoelastic fluid achieves a constant velocity, primarily parallel to the wall. This case is representative of the behavior for a wide range of $\theta_w$, excluding the limiting cases of $\theta_w \sim 90^\circ$ (near-negligible propulsion) and $\theta_w \sim 0^\circ$ (motion primarily away from wall). This symmetry-breaking is distinct from that experienced by a conventional low-$Re$ swimmer in Newtonian fluid, wherein a wall alters the trajectory of a swimmer or particle, but does not change the fundamental nature of propulsion\cite{Berke:2008dm,Lauga:2006ka,Lauga:2009p3877}; there, the particle can self-propel without the wall.

It important to note that in a Newtonian fluid, \emph{all} experiments discussed above yield negligible net displacement, comparable to the effects of $\vec F_\text{mag}$ and sedimentation when driving is turned off altogether (Fig.~\ref{fig_trajectories}[b, d]). Furthermore, the direction of displacement in Newtonian fluid is not controlled by particle shape or boundary conditions, confirming that it is not hydrodynamic in origin. Experiments with a symmetric dimer in a viscoelastic medium far from any boundaries also yield negligible net displacement (not shown).

\begin{figure}
\begin{center}
\includegraphics[width=3in]{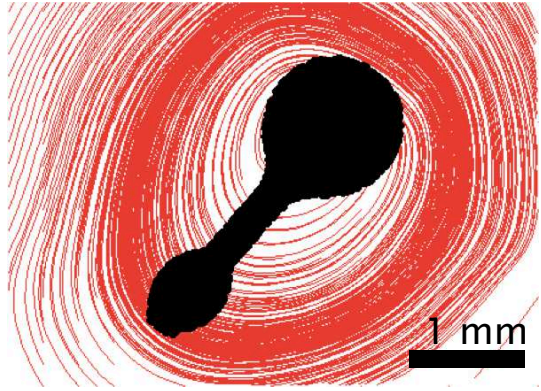}
\end{center}
\caption{(color online) Instantaneous streamlines, computed from particle tracking experiments, as a polar dimer rotates far from any boundaries, here plotted in the dimer frame. The different streamline curvature at each end suggests that forces due to the fluid's normal stress difference $N_1$ are unbalanced, contributing to propulsion.}
\label{fig_streamlines}
\end{figure}

Our experiments show that the net motion achieved by the dimers in the polymeric solution results from elastic stresses due to flow-induced changes in polymer conformation. These elastic stresses are history-dependent and do not entirely cancel out over one forcing period, but instead have a small rectified component that accumulates. A rheological property of polymeric solutions that is of particular relevance here is the first normal stress difference $N_{1}=\tau_{\theta\theta}-\tau_{rr}$, where $r$, $\theta$ and $z$ are cylindrical coordinates and $\tau$ is the fluid stress tensor. $N_{1}$ grows nonlinearly with fluid strain rate and, to lowest order, scales with strain rate as $\dot\gamma^2$,\cite{Bird:1987p2262} consistent with measurements of our own polymeric fluid.\cite{elmosupp} The combination of $N_{1}$ and curved streamlines in a given flow results in an inward-pointing volume force $-N_1 / r$ in the radial direction. In Fig.~\ref{fig_streamlines} we show instantaneous streamlines during the stroke of the polar dimer, computed from experimentally measured velocity fields. The curved streamlines around each bead, and the asymmetry in that curvature due to the dimer shape, strongly suggest that $N_1$ plays a role in propulsion. The greater strain rate and curvature at the small end of the dimer suggest that a stronger volume force there will move the dimer in the direction of its large end, consistent with observed propulsion.

\begin{figure}
\begin{center}
\includegraphics[width=6.0in]{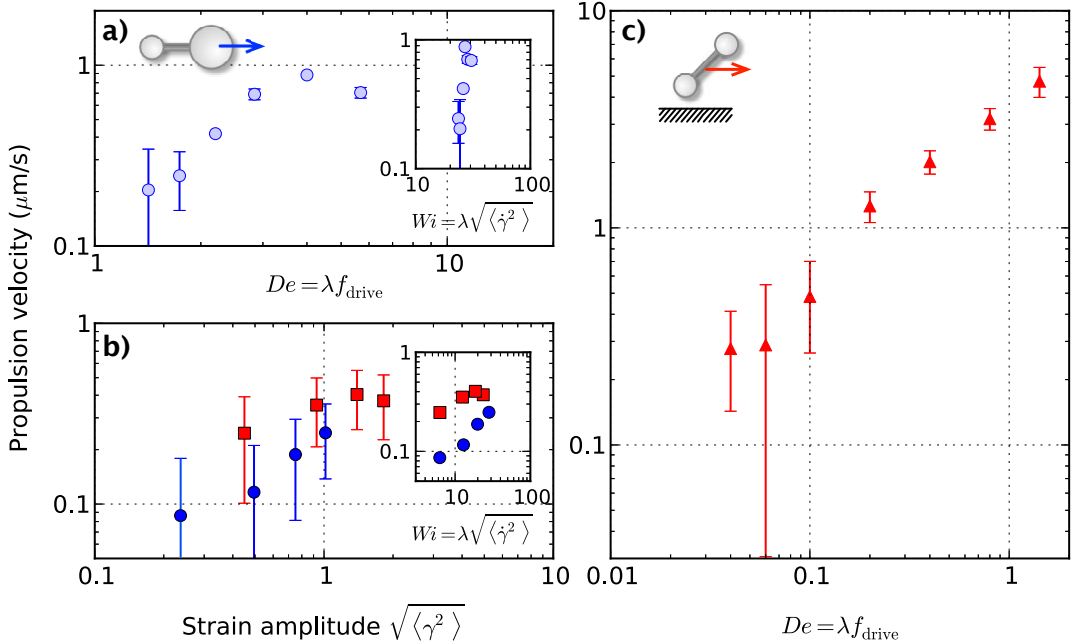}
\end{center}
\caption{(color online) Dependence of mean propulsion on driving parameters. Both types of particles show an increase in propulsion with Deborah number $De$, consistent with an elastic phenomenon. \textbf{(a)} Polar dimer. As $\fdrive$ is varied, propulsion appears to be controlled by $De$, not $\wi$ (inset), suggesting that the mechanism is not suited to a steady-flow description. Here $\dot \gamma \equiv \dot \theta \rdimer / \rbead$, where $\dot \theta$ is angular velocity. Note that strain amplitude decays as $\fdrive$ is increased, due to viscous resistance, accounting for the difference between the figure and its inset. This decay is the likely cause of the turnover at the highest $\fdrive$. \textbf{(b)} Polar dimer in a separate experiment (fluid viscosity $\sim 40\%$ higher) showing propulsion as a function of strain amplitude and $\wi$ (inset). Magnetic current amplitude is varied at $De = 2$ ($\fdrive = 0.7$~Hz, red squares) and $De = 4$ (1.4~Hz, blue circles). Data show a much weaker dependence on strain amplitude and $\wi$ than on $De$.
\textbf{(c)} Symmetric dimer at wall, showing much greater velocities and different scaling. Velocity component parallel to the wall is plotted. The range of $\fdrive$ is limited in order to maintain constant strain amplitude. Velocity measurements are cut off by noise at low $De$.}
\label{fig_scaling}
\end{figure}

To gain further insight into possible mechanisms, the effects of elasticity on propulsive speed are investigated for the polar and symmetric dimers, as shown in Fig.~\ref{fig_scaling}. The importance of elasticity is quantified by $De$. For both geometries, translation speed increases with elastic stresses for the range of $De$ investigated here. This trend is also seen for comparable $De$ in the cases of purely elastic propulsion numerically and theoretically investigated to date.\cite{Normand:2008p5115,Lauga:2009p8056,Fu:2009p3840,Pak:2010p5116}

As described above, the different streamline curvature at each end of the polar dimer suggests a role for normal stress difference effects as well as possible hydrodynamic interactions between the ends of the dimer.
However, a description of elastic effects based on steady shear, and thus on a single strain rate, is likely inadequate to model propulsion. This is seen in the insets of Fig.~\ref{fig_scaling}(a, b), where we plot velocity vs.\ Weissenberg number $\wi \equiv \lambda \dot \gamma$, using the characteristic strain rate $\sqrt{\langle \dot \gamma^2 \rangle}$. In experiments, we find that $\sqrt{\langle \dot \gamma^2 \rangle}$ is proportional to $\fdrive$ times strain amplitude. Propulsion is much more sensitive to $\fdrive$ than to strain amplitude, and so is poorly characterized by $\wi$. The dynamics would therefore best be modeled by fully accounting for the unsteadiness of the flow.

For the symmetric dimer results in Fig.~\ref{fig_scaling}(c), following the observations in Fig.~\ref{fig_trajectories}(c, d), we choose to plot the dominant velocity component, parallel to the wall, that is approximately constant over each movie. To keep geometry constant with respect to the fixed wall, we keep strain amplitude nearly constant by limiting $\fdrive$. In this geometry, elastic effects increase nearly linearly with $De$, unlike the roughly quadratic case of the polar dimer. We also find that even at these lower $\fdrive$, translation velocity is an order of magnitude greater than that of the polar case.

Propulsion of the symmetric dimer is inconsistent with two models we discuss here. First, while it is known that a particle moving steadily near a wall in viscoelastic fluid will experience a lift force away from the wall that scales as $\rbead / h$~(e.g. \cite{Hu:1999p6579}), Figs.~\ref{fig_trajectories}(c,~d) show that the dominant motion we observe is \emph{parallel} to the wall and decays little as $h$ increases. Second, Fu et al.\cite{Fu:2009p3840} show that for one type of swimmer in viscoelastic fluid, performing different parts of a reciprocal stroke at different rates enables propulsion. In our experiment, the dimer tip approaches the wall $\alt 10\%$ slower than on the return stroke, due to the nature of the magnetic driving. But on this basis, the cited analysis suggests net motion in the opposite direction of what we observe. The elastic mechanisms for this translation at a wall, and the propulsion of the polar dimer, thus remain open questions.

In summary, we have made the first experimental demonstration of propulsion by fluid elasticity, using reciprocally actuated dimers at low $Re$. The actuation yields no net motion in Newtonian fluid, but a small amount of polymer in the fluid adds an elastic response to the driving, permitting propulsion. The dimer shape directs propulsion in elastic fluid, without nearby boundaries. Near a wall, a dimer is propelled both parallel to and away from the boundary. All other propulsive strategies at low $Re$ in linear (e.g. Newtonian) fluids require a non-reciprocal stroke. Here, a swimmer may employ a reciprocal stroke provided the fluid is viscoelastic. We note that because time-reversibility is broken by the fluid and not the stroke, a time-reversal of driving does \emph{not} reverse propulsion, as it would for a non-reciprocal swimmer; here, the direction of propulsion is set by geometry alone.

We estimate propulsive (Froude) efficiency, defined as the fraction of total hydrodynamic power corresponding to net motion, to be $\mathcal{O}(1\%)$ --- comparable to the measured $\sim 2\%$ efficiency of the common low-$Re$ swimmer \textit{E. Coli}.\cite{chattopadhyay06}
The relatively slow propulsion we observe, at most $\sim 5$~$\mu$m/s, is due to the high fluid viscosity required to make sedimentation and shear-thinning effects negligible. We expect significantly faster propulsion in other realizations of this principle particularly for smaller dimers which would allow for viscosities that are much smaller. Also, fluid elasticity effects are expected to become more pronounced as the dimer (or swimmer) is miniaturized since the elasticity number, defined as $El=\lambda\mu/\rho L^{2}$, scales inversely with the square of the dimer length scale $L$. Finally, at higher $De$ or in geometries with greater streamline curvature, purely elastic instabilities may cause spontaneous propulsion without a wall or an asymmetric dimer shape --- or may greatly enhance the mechanisms demonstrated here.\cite{Larson:1990p8009,Shaqfeh:1996p8023,Pakdel:1996p7506} While organisms may not exploit the principle described here, viscoelastic media are common in nature,\cite{Lauga:2009p8056} and reliance on nonlinear rheology is not without precedent.\cite{denny80,lauga06}

Our work is also a proof-of-concept for an artificial ``swimmer'' that moves through complex fluid with only reciprocal actuation, a simple body shape, and no moving parts --- a less complicated design than for other propulsive strategies.\cite{Abbott:2009p2,Dreyfus:2005p1444} These principles could also be applied to pumps,\cite{Normand:2008p5115,Pak:2010p5116} or to exploiting other types of nonlinear fluid rheology. Further understanding of this effect and similar ones could greatly simplify fabrication of micro-swimmers for many artificial environments, or for biological settings where viscoelasticity is ubiquitous.

\begin{acknowledgments}
We thank
H.~Hu,
G.~Juarez,
M.~Moore,
and
A.~Morozov
for helpful discussions; and M.~Selman, X.N.~Shen, and G.~Friedman who contributed to early experiments. This work is supported by the Army Research Office through award W911NF-11-1-0488.
\end{acknowledgments}

\bibliography{references-nourl}

\end{document}